\documentclass[publish,JDR,paper,latex]{jdrarticle}
\usepackage[dvipdfmx]{graphicx}
\usepackage[dvipdfmx]{color}
\usepackage{tabularx}
\usepackage{url}
\usepackage{color}
\usepackage{ulem}
\usepackage{bm}
\usepackage{array}
\usepackage{amssymb}
\usepackage{ascmac}
\usepackage{mathtools}
\usepackage{amsmath,booktabs}
\usepackage{xcolor}
\usepackage{todonotes}
\usepackage{tabularx}
\usepackage{url}
\usepackage{color}
\usepackage{ulem}
\usepackage{bm}
\usepackage{array}
\usepackage{ascmac}
\usepackage{mathtools}
\usepackage{booktabs}
\usepackage{lineno}
\usepackage{lscape}
\usepackage{pxpgfmark}
\definecolor{brickred}{rgb}{0.8, 0.25, 0.33}
\DeclareRobustCommand{\remove}{\bgroup\markoverwith{\textcolor{red}{\rule[.5ex]{2pt}{0.4pt}}}\ULon}

\newcommand{\figref}[1]{\textbf{Fig.\ref{#1}}}
\newcommand{\listref}[1]{Table \ref{#1}}
\newcommand{\eref}[1]{Eqn. \eqref{#1}}
\newcommand{\secref}[1]{Section. \ref{#1}}

\newcommand*{\vertbar}{\rule[-1ex]{0.5pt}{5.5ex}}

\SetVolumeNumber{Vol.19 No.6, 2024} 
\SetArticleID{Dr24-0055; \jtoday}
\begin{document}

\pagestyle{jdrstyle}

\title{On the performance of sequential Bayesian update for database of diverse tsunami scenarios}
\author{Reika Nomura, Louise A. Hirao Vermare, Saneiki Fujita,  \\
Donsub Rim, Shuji Moriguchi, \\ Randall J. LeVeque and Kenjiro Terada}
\address{International Research Insititute of Disaster Science, Tohoku University, 468-1, Aramaki Aza Aoba, Aoba-ku, Sendai, 980-8579\\
E-mail: reika.nomura.a6@tohoku.ac.jp}
\markboth{R. Nomura, L. A. Hirao V., S. Fujita, D. Rim, S. Moriguchi, R. J. LeVeque and K. Terada}{On the performance of sequential Bayesian update for database of diverse tsunami scenarios}
\dates{00/00/00}{00/00/00}
\maketitle

\begin{abstract}
\noindent Abstract: 

Although the sequential tsunami scenario detection framework was validated in our previous work, several tasks remain to be resolved from a practical point of view.
This study aims to evaluate the performance of the previous tsunami scenario detection framework using a diverse database consisting of complex fault rupture patterns with heterogeneous slip distributions.
Specifically, we compare the effectiveness of scenario superposition to that of the previous most likely scenario detection method.
Additionally, how the length of the observation time window influences the accuracy of both methods is analyzed.
We utilize an existing database comprising 1771 tsunami scenarios targeting the city Westport (WA, U.S.), which includes synthetic wave height records and inundation distributions as the result of fault rupture in the Cascadia subduction zone.
The heterogeneous patterns of slips used in the database increase the diversity of the scenarios and thus make it a proper database for evaluating the performance of scenario superposition.
To assess the performance, we consider various observation time windows shorter than 15 minutes and divide the database into five testing and learning sets.
The evaluation accuracy of the maximum offshore wave, inundation depth, and its distribution is analyzed to examine the advantages of the scenario superposition method over the previous method.
We introduce the dynamic time warping (DTW) method as an additional benchmark and compare its results
to that of the Bayesian scenario detection method.
\end{abstract}

\begin{keywords}
Tsunami scenario detection, Synthetic database, Bayesian update
\end{keywords}

\section{Introduction}
Among the various types of natural disasters, earthquake-induced tsunamis are a common type \cite{sugawara2020trigger} and are particularly devastating to human society.
For instance, the 1960 Chilean earthquake generated waves ranging from 2 to 11 meters that hit the local community and propagated to Hawaii and Japan, reaching heights of over 5 meters \cite{cisternas2005predecessors}. The recurrence of an Mw 8.8 earthquake in 2010 also resulted in over 500 casualties due to the tsunami in Chile \cite{Franco_2010jdr}.
The 2004 Great Sumatra-Andaman Earthquake resulted in fatalities across 14 countries, with an estimated 70,000 casualties locally, ultimately reaching a total of 250,000 victims worldwide \cite{lay2005great}\cite{Tsuji_2006jdr}.

In recent decades, researchers have been increasingly drawn to the installed ocean network of observational gauges to mitigate such earthquake-induced tsunamis \cite{Takahashi_2017jdr}\cite{Matsumoto_2017jdr}\cite{Aoi_2019jdr}.
In countries situated around the Pacific Rim and Indian Ocean regions, sensors such as DART (deep-ocean assessment and reporting of tsunamis) buoys \cite{DART} have been deployed in near-shore and offshore areas.
The in-situ data provided by these sensors offer a more accurate reflection of tsunami characteristics than do seismic measurements (Bernard, 2005 \cite{bernard2015evolution}).
Furthermore, they are highly compatible with the ``database-type evaluation'' technique.
That is, in-situ data play a key role in searching for and detecting scenarios from a database when a tsunami occurs.
This enables us to complete time-consuming tsunami simulations prior to real events, with the only task remaining being to select the appropriate scenario for risk evaluation.
For instance, Titov et al. \cite{titov2005real} proposed an early tsunami warning system that employs a precomputed tsunami catalog.
With the recent focus on machine learning techniques (e.g., Fauzi and Mizutani 2019 \cite{fauzi2019machine}, Liu et al., 2021 \cite{liu2021comparison}; Mulia et al., 2022 \cite{mulia2022machine}, Kamiya et al. 2022 \cite{kamiya2022numerical}), database-type evaluations have emerged as a frontier topic.

In line with this context, we have developed a framework for sequential scenario detection based on unsupervised learning and Bayesian theory \cite{nomura2022sequential}.
In a previous study, we successfully detected scenarios with wave records very similar to those occurring within a seven-minute observation time window.
However, several practical issues still need to be addressed.

One critical aspect is the necessity of examining our method with a database consisting of diverse and complex fault rupture source patterns.
Geist (2022) \cite{geist2002complex} noted that heterogeneous slip pattern variations can significantly impact the ultimate tsunami simulation results in terms of wave amplitude, even when the earthquake moment, location, and geometry are identical.
Despite validating our framework with over six hundred tsunami synthetic scenarios propagated from various fault rupture locations \cite{koshimura_2022_zenodo}, we acknowledge that simple rectangular fault and homogeneous slip modeling can lead to each scenario being very similar.
Consequently, the database actually comprises a limited number of scenario clusters, where scenarios within each cluster share close information.
In such cases, the most likely scenario may indeed resemble the occurring event, as verified in the paper.
However, in the case of a diverse scenario database, the detected scenario may not necessarily exhibit similar tsunami characteristics, even if they are judged to be close according to Bayes' theorem.

To address this issue, synthesizing scenarios is a key concept.
In the previous paper, the potential of scenario superposition was not thoroughly addressed, but a brief report was provided.
In situations where a tsunami event is not closely aligned with any single scenario in the database, synthesizing candidate scenarios based on probability could be a feasible solution.
Additionally, exploring the balance between accuracy and observation time window length is essential for providing sufficient evacuation lead time.
Even though seven minutes was sufficient for selecting the proper scenario in the previous work, superposition techniques could achieve accurate risk prediction much faster.

With these remaining tasks in mind, this study aims to elucidate the performance of the previous framework of sequential Bayesian updates for tsunami scenario detection in a more practical context.
Specifically, in terms of the performance of probability-based scenario superposition, ``weighted mean averaging'' type detection is investigated with a tsunami scenario database that comprises diverse fault rupture patterns.
First, the developed framework for tsunami scenario detection is briefly introduced.
After introducing the proper orthogonal decomposition and sequential Bayesian update techniques, we also introduce weighted mean averaging scenario detection and the previous most likely tsunami scenario detection methods.
The well-known dynamic time warping (DTW) distance measure is used as the benchmark that assesses the advantages of these two Bayesian update-based evaluation methods.
Subsequently, we introduce a tsunami scenario database focusing on the city of Westport in Washington (U.S.) \cite{williamson2020source}.
By investigating the arrival time and maximum height trends at a nearshore gauge in different scenarios, we establish various observation time window conditions and split the data into 5 different sets of testing/training data for cross-validation.
Based on the 5-fold cross-validation results for each method, we discuss the most reasonable observation time windows for various cases.
In addition to the questions regarding the observation window size, we explore the possibility of improving predictions based on weighted mean averaging.
For both scenario detection and weighted mean averaging, we investigate the accuracy of maximum offshore wave height, maximum offshore inundation depth, and inundation zone area predictions.

The Westport Peninsula is used as a test case due to the availability of good training and testing data from previous studies and because it is a challenging test case for real-time forecasting, due to the short time between the earthquake and the tsunami arrival.  But we note that in practice the ground motion itself would be the primary warning for a CSZ tsunami in this particular community, and public education of tsunami hazards is a critical component for life safety.

\section{Sequential tsunami scenario detection}
This section briefly introduces the tsunami scenario detection framework based on both proper orthogonal decomposition (POD) and sequential Bayesian updating.
The framework used in this study consists of two steps: a preprocessing step, which involves constructing the database, and a real-time step, where scenarios most similar to the occurring event are identified.

In the following section, we briefly remind readers of the POD method for extracting characteristics of each tsunami scenario, as well as the Bayesian update process that utilizes the extracted POD features to evaluate the scenario probability from the in situ observational data.

We do not delve into the details of each process in this section.
For a comprehensive explanation of the entire process, readers are referred to Nomura et al. \cite{nomura2022sequential} or Fujita et al. \cite{fujita2024optimization}. For further information on POD, please refer to Liang et al. \cite{liang2002proper} or Kerschen et al. \cite{kerschen2005method}.

\subsection{POD for extracting the tsunami characteristics}
\label{sec:POD}
First, we assume that we have $N_s$ tsunami scenario databases consisting of the wave data at all $N_g$ gauges for $N_t$ time steps (with a total data size of $N_s \times N_t \times N_g$).
According to the rule introduced in a previous paper \cite{nomura2022sequential}, we can establish the data matrix $\bm{X} \in \textrm{I\!R}^{N_g \times (N_t\times N_s)}$ by combining all the wave history data $\bm{\eta}_j^{(t_m)}$ ($j=1, \ldots N_s$, $m=1,\ldots N_t$).
Here, the data matrix $\bm{X}$ can be alternatively expressed as a reduced order data matrix $\tilde{\bm{X}} \in \textrm{I\!R}^{N_g \times (N_t\times N_s)}$ as follows:
\begin{equation}
    \bm{X} \approx \tilde{\bm{X}} = \bm{\Phi}_r \bm{A},
    \label{eq:X_dcomp}
\end{equation}
where the low-rank mode matrix $\bm{\Phi}_{r} \in \textrm{I\!R}^{N_g \times r}$ and the coefficient matrix $\bm{A} \in \textrm{I\!R}^{r \times (N_t\times N_s)}$ are rewritten as
\begin{align}
\bm{A}
  &=
\begin{bmatrix}
 \bm{\alpha}_{1}   & \bm{\alpha}_{2} & \ldots & \bm{\alpha}_{j} & \ldots &\bm{\alpha}_{N_s}  \\
\end{bmatrix},
\quad
\notag \\
  \text{with} &
\quad
\bm{\alpha}_{j} = 
\begin{bmatrix}
 \vertbar                   & \vertbar                  &       & \vertbar \\
 \bm{\alpha}_{j}^{(t_1)}    & \bm{\alpha}_{j}^{(t_2)}   & \ldots& \bm{\alpha}_{j}^{(t_m)} \\
 \vertbar                   & \vertbar                  &       & \vertbar
\end{bmatrix},
  \label{eq:each_matrix2}
\end{align}
Here, $\bm{\alpha}_{j}^{(t)} \in \textrm{I\!R}^{r \times N_t}$ contains a series of coefficients corresponding to scenario $j$ at time $t$ and is used in the subsequent Bayesian update steps.

To make $\tilde{\bm{X}}$ a good approximation of the original data $\bm{X}$ ($\|\bm{X}-\tilde{\bm{X}}\| < \epsilon$), the following cumulative contribution ratio is employed to determine the number of required modes $r$ in \eqref{eq:X_dcomp}:
\begin{equation}
\textrm{c}(r) = \dfrac{\sum_{j=1}^{r} \lambda_{j}}{\sum_{j=1}^{N_g} \lambda_{j}}.
 \label{eq:contribution}
\end{equation}

\subsection{Bayesian update}
\label{sec:bayesian_update}

Assuming that an unknown tsunami event occurs and that its propagating waves are sequentially observed as $\bm{\eta}_{\chi}^{t}$, the current tsunami event-specific feature $\tilde{\bm{\alpha}}_{\chi}^{t}$ can be extracted by the inverse formula of \eqref{eq:X_dcomp} as
\begin{equation}
    \tilde{\bm{\alpha}}_{\chi}^{(t)} = \bm{\Phi}_r^{\dagger} \bm{\eta}_{\chi}^{(t)}, 
    \label{eq:pinverse}
\end{equation}
where $\bm{\Phi}_r^{\dagger}$ is a pseudoinverse of $\bm{\Phi}_r$. We use the Moore--Penrose inverse for the actual calculation as well as in the early study.
We use the current scenario specific data $\bm{\alpha}_{\chi}^{(t_m)}$ as the ``key'' and the coefficient matrix $\bm{A}$ as the ``database''.
In the following, we evaluate the similarity of the current event to the $j$-th scenario in the database:
\begin{equation}
   \bm{\Delta}_{j}^{(t)} =
  \left\{ \left(\bm{\alpha}_j^{(t)} - \tilde{\bm{\alpha}}_{\chi}^{(t)} \right)^{T} \bm{P}^{t^{-1}} 
  \left(\bm{\alpha}_j^{(t)} - \tilde{\bm{\alpha}}_{\chi}^{(t)} \right) \right\}^{1/2},
  \label{eq:Mahalanobis}
\end{equation}
where $\bm{\Delta}_j^{t}$ is the Mahalanobis distance, which reflects the similarity between two events $\bm{\alpha}_j^{(t_m)}$ and $\tilde{\bm{\alpha}}_{\chi}^{(t_m)}$ based on the results of POD, and $\bm{P}^{t}$ is the covariance matrix, which is defined as $\bm{P}^{t} = 0.1 \bm{\Sigma}^{1/2}$, where $\bm{\Sigma}$ is the degree of contribution of each spatial mode.

The defined Mahalanobis distance \eqref{eq:Mahalanobis} is then used as the measure of similarity in the Bayesian formulation:
\begin{equation}
  P \left(E_{j}^{t_m} \mid \varepsilon^{t_m}  \right) =
  \frac{ L\left( E_{j} \right)} 
  {\sum_{i=1}^{N_{s}}  L\left( E_{i} \right)  \left(E_{i}^{t_{m-1}} \right)}
  P \left(E_{j}^{t_{m-1}} \mid \varepsilon^{t_{m-1}}  \right), 
  \label{eq:bayesian_update}
\end{equation}
\begin{equation}
  L\left( E_{j} \right) =
  \frac{1}{\sqrt{\left( 2\pi \right)^{r} \mid \bm{P}^{t}\mid} }
  \exp \left(-\frac{1}{2}  {\left(\bm{\Delta}_{j}^{t}\right)} ^{2} \right),
  \label{eq:like_function}
\end{equation}
where $L\left( E_{j} \right)$ is the likelihood of the $j$-th scenario, denoted by $E_j^{t_m} (j=1, 2, \ldots N_s)$, being selected at the current time $t$, which is updated at each time step with the observational measure $\varepsilon^{t_m}$.

\subsection{Tsunami scenario detection}
\label{sec:3method}
We first define the tsunami risk indices that we finally want to obtain as the results of the scenario detection.
As mentioned in Sec. \ref{sec:POD}, we have wave history data $\bm{\eta}_{j}^{(t_m)}$ for all the simulation times ($t = [t_0, t_{N_t}]$); thus, we can evaluate the following maximum wave height $\eta_{n',j}^{\max}$ as the index for determining the tsunami risk:
\begin{equation}
    \eta_{n',j}^{\max} =  \max_{t}{\eta_{n',j}^{t}}
\end{equation}
where $n'$ is the number of targeted gauges.
In addition, we assume that we have the inundation risk indices $\bm{h}_{\max}$ and $H_{\max}$ as a result of the precomputed tsunami scenario simulation.
The former index $\bm{h}_{\max}^{j}$ represents the $n_x \times n_y$ matrix for storing the maximum inundation depth distribution during event $j$:
\begin{equation}
    \bm{h}_{\max}^{j} = \begin{bmatrix}
        h_{\max}^j(x_1, y_1) &  \hdots & h_{\max}^j(x_1, y_{n_y}) \\
        h_{\max}^j(x_2, y_1) &  \hdots & h_{\max}^j(x_2, y_{n_y}) \\
        \vdots  & \ddots & \vdots \\ 
        h_{\max}^j(x_{n_x}, y_1) & \hdots & h_{\max}^j(x_{n_x}, y_{n_y}) \\
    \end{bmatrix}
    \label{eq:h_max_dist}
\end{equation}
The latter index $H_{\max}^j$ is a scalar value that represents the maximum inundation depth among all the targeted domains:
\begin{equation}
   H_{\max}^{j} = \max{\bm{h}_{\max}^{j}}.
   \label{eq:H_max}
\end{equation}

When an unknown tsunami event $\chi$ occurs, we can predict the offshore wave height $\eta_{n',\chi}^{\max}$, maximum inundation distribution $\bm{h}_{\max}^{\chi}$ and maximum inundation depth $H_{\max}^\chi$ by utilizing the results of Bayesian updating.
In other words, we can either select the scenario with the highest conditional probability as the ``most probable scenario'' or obtain a ``weighted mean average'' by synthesizing scenarios based on their probabilities.
The former method has been presented in a previous paper; therefore, the main focus of this study is to examine the advantages of the latter method over the former.

In addition to comparing the two methods, we also need to address the primary question of whether they both perform effectively on a scenario database comprising diverse fault rupture patterns.
For this purpose, a nonprobabilistic measure that can quantitatively evaluate the similarity of wave history data is also needed.
Although there are several indices available for judging the matching of two sets of time series data (e.g., Yamamoto et al., 2016 \cite{yamamoto2016multi}), this study employs dynamic time warping (DTW) \cite{bellman1959adaptive} for the benchmark.
As mentioned in Senin 2008 \cite{senin2008dynamic}, DTW is a method designed to detect similar time series data while minimizing the effects of shifting and distortion. 
It can measure the closeness of the time series with similar waveforms and amplitudes but different phases, making it generally more robust than general Euclidean distance-based comparison \cite{lee2020clustering} \cite{ouyang2010similarity}.

\subsubsection{Most probable scenario}
As introduced in our earlier work \cite{nomura2022sequential}, the scenarios with the highest probabilities $J$ are identified as the scenarios most similar to the occurring event,
\begin{equation}
        J = \textrm{arg} \max_{j} \ P(E_{j}^{t_{obs}} \mid \varepsilon^{t_\textrm{obs}}).
        \label{eq:most_scenario}
    \end{equation}
Thus, we can use the data from scenario $J$ in our predictions as follows:
\begin{align}
       \eta_{n',\chi}^{\max} = \eta_{n',J}^{\max}, \label{eq:eta_most} \\ 
      \bm{h}_{\max}^{\chi} = \bm{h}_{\max}^{J}, \label{eq:hmax_most} \\
      H_{\max}^{\chi} = H_{\max}^{J}. \label{eq:hmax_all_most}
\end{align}

\subsubsection{Weighted mean scenario}
In this case, predictions are obtained based on the superposition of data from every scenario considering the conditional probability, as follows:
\begin{align}
      \eta_{n',\chi}^{\max}  = \sum_{j=1}^{N_s} P(E_{j}^{t_{obs}}\mid \varepsilon^{t_{obs}})\cdot \eta_{n',j}^{\max}, \label{eq:eta_mean} \\
      \bm{h}^{\chi}_{\max}  =  \sum_{j=1}^{N_s} P(E_{j}^{t_{obs}}\mid \varepsilon^{t_{obs}})\cdot \bm{h}^{j}_{\max}, \label{eq:hmax_mean} \\
      H_{\max}^{\chi} =\sum_{j=1}^{N_s} P(E_{j}^{t_{obs}}\mid \varepsilon^{t_{obs}})\cdot H_{\max}^{j}.
       \label{eq:hmax_all_mean}
    \end{align}

\subsubsection{Shortest DTW distance scenario}
Similar to the most likely scenario detection method, in other words, selecting the scenario with the shortest Mahalanobis distance, as described in Equation \ref{eq:most_scenario}, we can detect the scenario $j^*$ that has the shortest DTW distance as follows:
\begin{equation}
        j^{*} = \underset{j \in 1,2,\dots N_s} {\operatorname{argmin}} d_{\chi, j}^{[t_0, t_m]}
        \label{eq:DTW_scenario}
    \end{equation}
where $d_{\chi, j}^{[t_0, t_m]}$ represents the DTW distance set calculated from the wave history data for scenarios $j$ and $\chi$ within the time window [$t_0$, $t_m$].
Here, the risk prediction can be performed as follows:
\begin{align}
       \eta_{n',\chi}^{\max} = \eta_{n',j^*}^{\max}, \label{eq:eta_dtw} \\ 
      \bm{h}_{\max}^{\chi} = \bm{h}_{\max}^{j^*}, \label{eq:hmax_dtw} \\
      H_{\max}^{\chi} = H_{\max}^{j^*} \label{eq:hmax_all_dtw}.
\end{align}

\section{Diverse tsunami scenario database} \label{CSZdata}

\subsection{Synthetic tsunami scenario database comprising diverse and complex fault rupture patterns}
\begin{figure*}[!ht]
    \centering
    \includegraphics[width=0.8\linewidth]{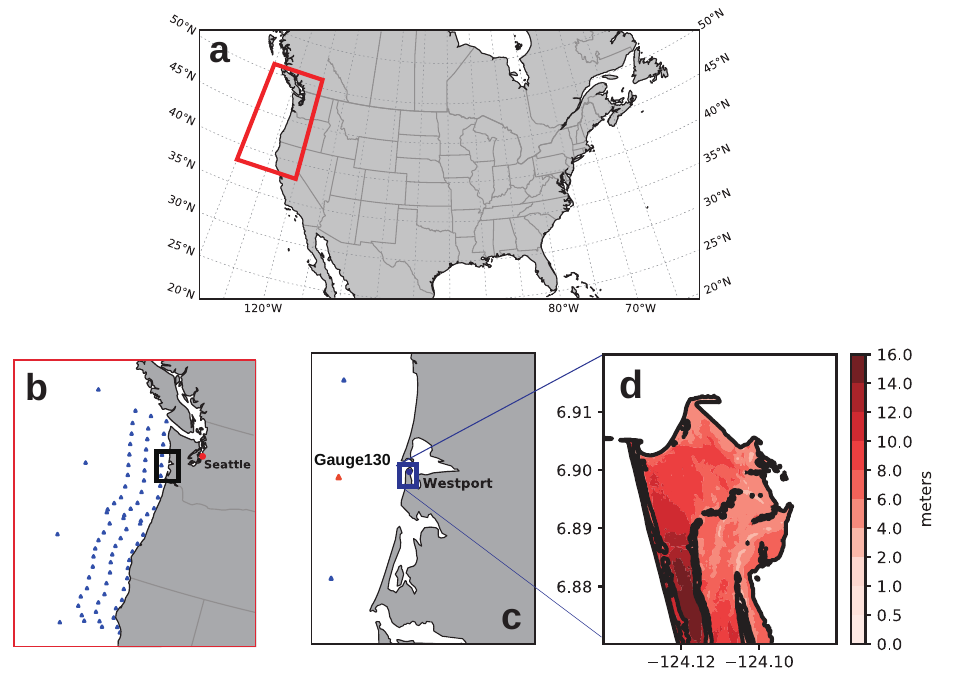}
\caption{Location of the city of Westport and the observation gauges. (a-c) Westport is a coastal city located in Washington state, approximately 100 km from the Cascadia subduction zone. (d) Topography elevation on the Westport Peninsula.}
    \label{fig:Westport_gauges}
\end{figure*}

As introduced in the first section, a scenario database consisting of diverse fault rupture patterns is required to determine the robustness of the framework.
For this, an established tsunami scenario database for Westport, which is located in Washington state, U.S., as illustrated in \figref{fig:Westport_gauges}\textbf{a}, \textbf{b}, and \textbf{c}, was employed.
Westport is a coastal city exposed to tsunami risk caused by earthquakes arising from the Cascadia subduction zone (CSZ).
The CSZ is known to generate earthquakes with magnitudes larger than Mw8.0 at a return period of approximately 500 years \cite{thomas1987earthquake}.
The last event was the 1700 Cascadia earthquake, which was an M8.7 $\sim$ 9.2 earthquake that induced a tsunami \cite{atwater2011orphan}.
The CSZ can generate major earthquakes that threaten communities along the northwestern coast of North America \cite{goldfinger2003holocene}.

\begin{figure*}
   \centering
    \includegraphics[scale=0.9]{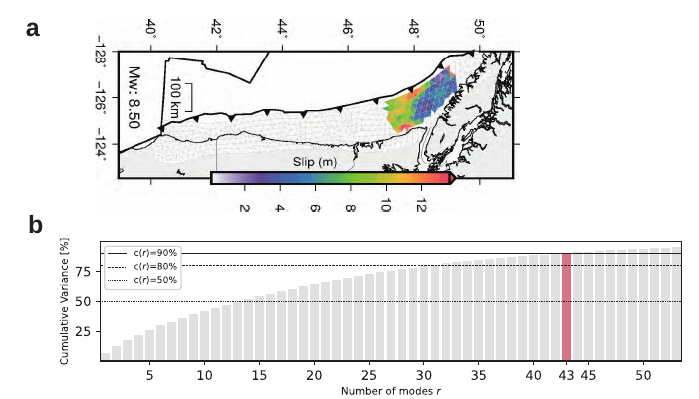}
\vspace{-3 mm}
\caption{\textbf{a} 
The CSZ fault region with a sample heterogeneous slip distribution (from the open repository \cite{UW_repo} that summarizes the data used in Williamson et al. 2020 \cite{williamson2020source}), \textbf{b} The cumulative contribution $\textrm{c}(r)$ defined in \eref{eq:contribution}.}
\label{fig:fault_model}
\end{figure*}

We used the same database as that generated in Williamson et al. 2022\cite{williamson2020source}.
They first prepared 2000 slip distributions with defined earthquake magnitudes of M7.5, M8.0, M8.5 and M9.0.
With the Karhunen-Lo\'eve (KL) expansion \cite{leveque2016generating} with triangular subfaults implemented in the \texttt{fakequake} \cite{melgar2016kinematic} module of the \texttt{MudPy} software package \cite{melgar2021mudpy}, various random lognormal slip distributions were statistically generated.
Here, \figref{fig:fault_model}\textbf{a} shows one of the fault rupture realizations generated by Williamson \cite{williamson2020source}.
Comparing the rectangular fault model, illustrated in \figref{fig:fault_model}\textbf{b}, with the uniform slip distribution model (e.g., \cite{koshimura_2022_zenodo}, \cite{koshimura_2023_zenodo}), each slip distribution is heterogeneous in the current database, as represented by the contour colors in \figref{fig:fault_model}\textbf{b}.
Thus, the resultant tsunami scenarios should be more diverse than those of rectangular fault slip models.

For the tsunami simulation results, the 4-hour wave height data sampled at 76 synthetic gauges located along the coast, as illustrated in \figref{fig:Westport_gauges}\textbf{b}, were stored.
Here, the nearest gauge, ``Gauge 130'', which is illustrated as a red triangle in \figref{fig:Westport_gauges}\textbf{c}, is approximately 30 km away from the coast.
The time between each wave snapshot varies since GeoClaw \cite{clawpack,berger2011geoclaw,mandli2016clawpack}, a tsunami simulation tool, uses an adaptive mesh refinement method.
Therefore, the wave history data were interpolated to 0.2 Hz, i.e., 5-second increment data in this study.
Thus, $\{ t_1, t_2, \cdots t_{N_t} \}$ is equal to $\{5.0, 10.0, \cdots (60[\textrm{sec} /\textrm{min}] \cdot 60 [\textrm{min}/\textrm{hr}] \cdot 2 [\textrm{hr}] )  \}$; see \cite{williamson2020source} for more details.

In addition to the offshore tsunami wave data, the onshore inundation depth distribution $\bm{h}_{\max}^{j}$ of each scenario was obtained at 216 $\times$ 180 locations in the Westport area ($\bm{h}_{\max} \in \textrm{I\!R}^{216 \times 180}$), as shown in \figref{fig:Westport_gauges}\textbf{e}.
The southwest and northeast edge points of this area are ($x_1$, $y_1$) = (W$124^{\circ} 4'48.50"$, N$46^\circ52'12.50"$) and
($x_{216}$, $y_{180}$) = (W$124^\circ 8'23.50"$, N$46^\circ 55'11.50"$), with a 1 second interval.
In addition, the maximum $h^{j}_{\max}(x, y)$ values among all the 216 $\times$ 180 = 3880 observation points are also defined according to \eref{eq:H_max}.

To confirm the diversity of the scenario database, we examined the number of modes $r$ in \eref{eq:X_dcomp} and \eqref{eq:contribution}.
Here, a small number of modes $r$ implies that the database can be represented by low-rank information, indicating that each scenario can be almost entirely expressed by common information shared across other scenarios in the database.
Therefore, a diverse scenario database would require more modes to reconstruct the original data with minimal information loss.
\figref{fig:fault_model}\textbf{c} illustrates the relationship between $r$ and c$(r)$ resulting from POD analysis of the wave history data $\eta_{j}^{(t_m)}$.
The pink bar indicates that 43 modes, which is more than half of the total number of modes (43/76), are necessary to reconstruct 90\% of the original data matrix $\bm{X}$.
This suggests that the current database exhibits a very complex structure due to the intricate pattern of fault slip.


\subsection{Observation time window setup}
\begin{figure*}[!htb]
   \centering
    \includegraphics[width=0.8\linewidth]{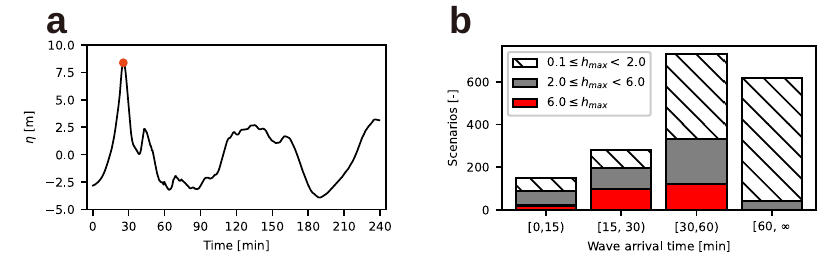}
\vspace{-3 mm}
\caption{\textbf{a} Wave history data and its wave arrival time $t_{arrv}$. \textbf{b} Histograms of the scenarios based on the first wave arrival time.
The colors and hatching denote the magnitude of the local maximum inundation depth.
$H_{\max}$.}
    \label{fig:hist_database}
\end{figure*}

To identify a reasonable observation time window length $t_{obs}$ in relation to the prediction accuracy, we first analyzed each scenario in the database in terms of the offshore maximum wave heights $\eta_{n', j}^{\max}$, maximum inundation depth $H_{\max}^{j}$ and wave arrival time $t_{arrv}$.
The tsunami arrival time $t_{arrv}$ and maximum wave height were both defined at the gauge nearest to the coast of Westport, namely, Gauge 130.
As illustrated in \figref{fig:hist_database}\textbf{a} as a red point, the time of the first wave peak was recorded as the arrival time $t_{arrv}$.
The maximum wave height $\eta_{n',j}^{\max}$ was recorded at the first wave arrival time $t_{arrv}$ in this case. All $t_{arrv}$ were identified using the \texttt{find\_peaks} package of the \texttt{scipy} \texttt{python} library.

We first eliminated the scenarios with only small wave amplitudes at Gauge 130.
As a result of rejecting scenarios with threshold values $\eta^{\max}_{n',j} < 0.01$, the total number of scenarios $N_s$ we used in this study was 1771, as summarized in \listref{tab:job_cases}.

\figref{fig:hist_database}\textbf{b} shows a histogram summarizing the number of scenarios classified according to the wave arrival time $t_{arrv}$.
The red and gray zones represent the scenarios with relatively high inundation risks of $2.0 \leq H_{\max}^{j} < 6.0$ and $6.0 \leq H_{\max}^{j}$, respectively.
As depicted in the histograms, scenarios classified as being of high inundation risk are those with tsunamis arriving within an hour.
Specifically, the greatest total amount of inundated water in the high-inundation-risk scenarios occurs within the interval of $15 \textrm{min} \leq t_{arrv} < 30 \textrm{min}$.
Assuming that waves need several tens of minutes to reach the coast of Westport after passing through Gauge 130, we set up 10 different $t_{obs}$ conditions shorter than 15 minutes, namely, $t_{obs} = 1~\textrm{min}, 2~\textrm{min}, \ldots, 15~\textrm{min}$, as candidates for reasonable observation time windows. Table 1 summarizes these conditions as well as other specific information about the database.

Based on the general concept of k-fold cross-validation, we first generated 5 pairs of 1417 learning scenarios and 354 test scenarios, accounting for 80\% and 20\%, respectively, of the whole dataset to avoid data sampling bias.
Each scenario in the test data group was treated as a real-time event $\chi$.
Additionally, the wave history data for the 1417 learning scenarios were used in POD and the Bayesian updating process.
Therefore, $N_s$ was equal to 1417.
With POD, we extracted 71 coefficient matrices $\bm{\alpha}_{j}^{(t_m)}$ ($j=1, 2, \ldots 1417, m= 1, 2, \ldots 2880$) at each time step.
From the wave data for each test scenario, we evaluated the coefficient matrices $\tilde{\bm{\alpha}}_{\chi}^{(t_m)}$ via an inverse calculation approach.

\begin{table*}
    \centering
    \caption{Conditions of 5-fold cross validation}
    \footnotesize
    \begin{tabular}{l|r}
    \toprule
       Total number of scenarios & 1771 \\
       (Ratio of the numbers of training/testing scenarios) & 1417 : 354 \\
       Number of gauges $N_g$ & 76 \\
       Observation time window $t_{obs}$ & 1~min, 2~min, 3~min, 4~min, 5~min, 6~min,\\ 
       &8~min, 10~min, 12~min, 15~min\\
    \bottomrule
    \end{tabular}
    \label{tab:job_cases}
\end{table*}

\section{Results of scenario detection with various observation time windows}
\label{sec:results}
As mentioned in the previous section, fivefold cross-validation was carried out.
In each fold, i.e., for each pair of test/training scenarios, both scenario detection methods introduced in Sec.\ref{sec:3method}, namely, most probable and weighted mean, were applied.
The accuracy of the predicted risk indices, namely, the maximum wave height $\eta_{n',\chi}^{\max}$, maximum inundation distribution $\bm{h}_{\max}^{\chi}$ and maximum inundation depth $H_{\max}^{\chi}$, were then investigated to evaluate the advantage of weighted mean averaging-type scenario detection over the previous most likely scenario detection.
DTW distance-based predictions were supplementary used to evaluate the prediction performance at each observation time window.

The performance of weighted mean scenario detection in each observation time window was evaluated according to the prediction accuracy of each of the three indices, as described in the following.
\subsection{Wave history data $\eta^{\max}_{n',\chi}$}
\begin{figure*}[!ht]
  \centering
  \includegraphics[width=0.8\linewidth]{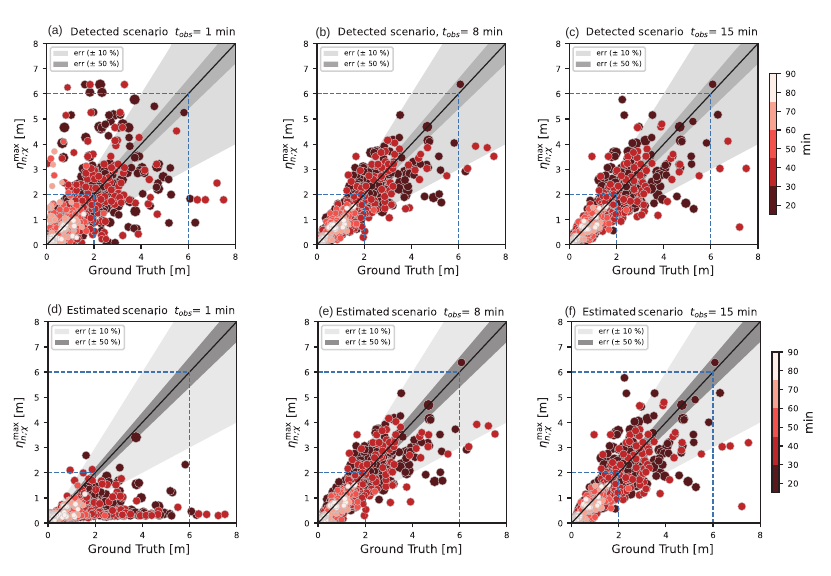}
\caption{The results of maximum wave height $\eta_{n',\chi}^{\max}$ detection. (a), (b), and (c) show the predictions based on the most likely scenario \eqref{eq:eta_most} for observation time windows of $t_{obs}$ =1~min, 8~min, and 15~min, respectively.
(d), (e), and (f) show the predictions based on the weighted mean scenario \eqref{eq:eta_mean} for observation time windows of $t_{obs}$ =1~min, 8~min, and 15~min, respectively.}
  \label{fig:etamax_scatter}
\end{figure*}

The scatter plots in \figref{fig:etamax_scatter} show the relationship between the forecasted $\eta_{n', \max}^{\chi}$ and ground-truth values at Gauge 130 (\figref{fig:Westport_gauges}\textbf{c}).
\figref{fig:etamax_scatter}(a)-(c) shows the prediction based on the detected most likely scenario (\eref{eq:eta_most}), and \figref{fig:etamax_scatter}(d)-(f) shows the prediction based on the weighted mean of all the learned scenarios (\eref{eq:eta_mean}).
There are 1770 (354 $\times 5$) circles plotted in each panel, reflecting the evaluations for Gauge 130 (shown in \figref{fig:Westport_gauges}) based on 5-fold cross-validation.
The color and size of each circle indicate the magnitude of the wave arrival time $t_\textrm{arrv}$.

In panels (b) and (e), which represent $t_\textrm{obs}= 8~\textrm{min}$ in the middle column, the numbers of circles plotted outside the gray zone, corresponding to 50 or 10\% error in reference to the ground truth, are smaller than those in panels (a) and (d).
However, the accuracies do not seem to change drastically for $t_{obs}=15~\textrm{min}$.
We focus on these errors in terms of the relation with the wave arrival time $t_{arrv}$.
The fast-arriving waves, which usually correspond to large wave heights, are more drastically over- or underestimated than the late-arriving waves, represented by white/pale circles.
Many of these trends are unresolved, even if the observation time window $t_{obs}$ is extended to 15 min.

Additionally, these results suggest that scenario estimation based on the weighted mean does not necessarily improve accuracy.
As shown in \figref{fig:etamax_scatter}(f), the $\eta_{n',\chi}^{\max}$ values estimated from the weighted mean operations are not very different from those plotted in panel (c), which illustrates the results of scenario detection.
Moreover, the weighted mean operation can easily result in the significant underestimation of wave heights if the observation time window $t_\textrm{obs}$ is too short, as shown in \figref{fig:etamax_scatter}(d).

\begin{figure*}[!ht]
    \centering
    \includegraphics[width=0.8\linewidth]{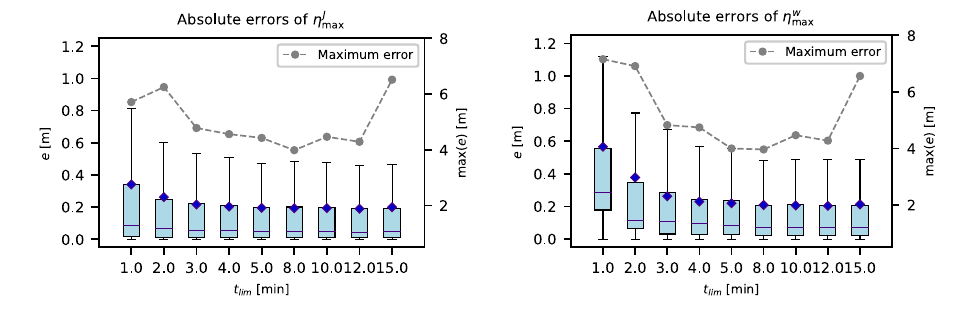}
\caption{The variance in the maximum wave height prediction error at Gauge 130 during each observation period $t_{obs}$. (a) Results based on most likely scenario detection, and (b) results based on the weighted mean scenario. The diamond plot, horizontal line inside the box, and edge lines of the whiskers represent the mean, median, and min/max values, respectively.}
    \label{fig:etamax_sigma_box}
\end{figure*}

These tendencies are further summarized in the box plots shown in \figref{fig:etamax_sigma_box}.
The two box plots illustrate the variance in the absolute error $e$ of (a) $\eta_{n',\chi}^{\max}$ based on most likely scenario detection with \eqref{eq:eta_most} and (b) $\eta_{n',\chi}^{\max}$ based on weighted mean estimation with \eqref{eq:eta_mean}.
The diamond plot, horizontal line inside the box, and edge lines of the whiskers represent the mean, median, and min/max values, respectively.
In both panels, the worst prediction accuracy is observed for the shortest observation time window $t_\textrm{obs} = 1~\textrm{min}$ based on both the mean values (diamond plots) and the upper whiskers.
As the length of the observation time window increases to $t_\textrm{obs}=2~\textrm{min}, 3~\textrm{min}$, the absolute error decreases.
However, the improvements generally stabilize after eight minutes.
This tendency is common for both prediction methods, and the error variances are comparable between them.

To determine whether these absolute errors are acceptable, we compare the scenario detection results based on DTW distance measures.
The leftmost boxes in panels \figref{fig:etamax_sigma_box} (a) and (b) represent the $\eta^{\max}_{\chi}$ predictions based on the shortest DTW distance scenario $j^*$, as defined based on \eref{eq:DTW_scenario}.
The DTW distances are all calculated by the \texttt{fastdtw} module provided in the scripting language \texttt{Python}.
Focusing on the interquartile range (IQR), which is equivalent to the vertical length of the box, the predictions based on both the most likely scenario and the weighted mean scenario are comparable or superior to those based on the DTW method if we set $t_{obs}$ to a value longer than several minutes.


\subsection{Maximum inundation depth $H_{\max}$}
\begin{figure*}
    \centering
    \includegraphics[width=0.8\linewidth]{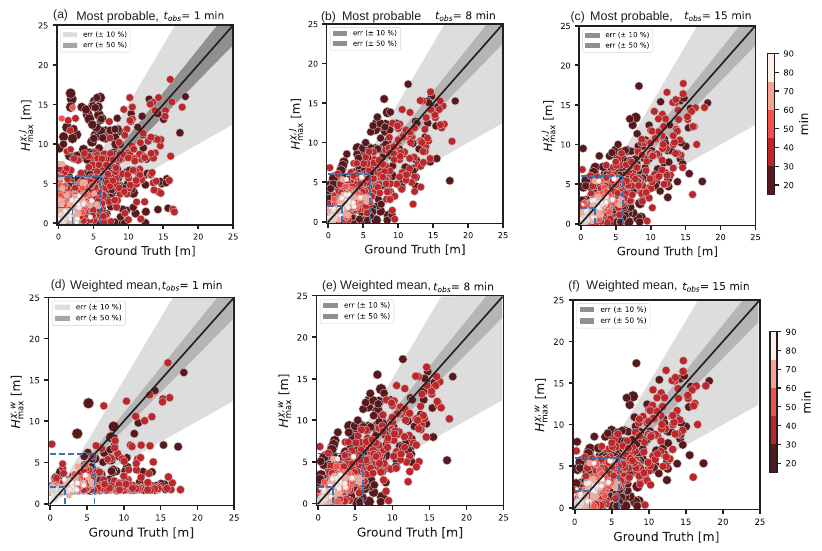}
\caption{The maximum inundation depth $H_{\max}^{\chi}$. (a), (b), and (c) show the predictions based on the most likely scenario method \eqref{eq:hmax_most} for observation time windows of $t_{obs}$ =1~min, 8~min, and 15~min, respectively.
(d), (e), and (f) show the predictions based on the weighted mean scenario method \eqref{eq:hmax_mean} for observation time windows of $t_{obs}$ =1~min, 8~min, and 15~min, respectively.}
    \label{fig:hmax_scatter}
\vspace{3 mm}
    \centering
    \includegraphics[width=0.8\linewidth]{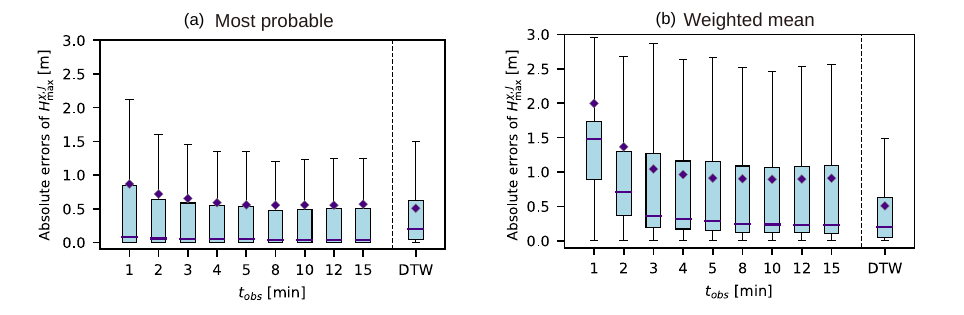}
\vspace{-6 mm}
\caption{The variance in the inundation errors during each observation period $t_{obs}$. (a) Results based on most likely scenario detection, and (b) results based on the weighted mean scenario method. The diamond plot, horizontal line inside the box, and edge lines of the whiskers represent the mean, median, and min/max values, respectively.}
    \label{fig:hmax_boxplot}
\end{figure*}

The same tendencies as those of the wave height predictions were observed in the evaluation of $H_{\max}^{\chi}$, as shown in \figref{fig:hmax_scatter}.
The upper row shows the results based on most likely scenario detection (\eref{eq:hmax_most}), and the lower row shows the results based on the weighted mean scenario (\eref{eq:hmax_mean}).
The circles are not close to the centerline, which would be indicative of perfect accuracy.
In particular, significant under/overestimations occur in the rapid-arrival tsunami cases regardless of the inundation depth.
This must be because the Bayesian updates are based on the modes and coefficient matrices extracted from the propagated wave information.
Thus, the inundation depth, which is not directly used in probability evaluations, is slightly harder to precisely predict than the wave height.

Regarding the two methods of prediction defined in \eref{eq:hmax_most} and \eref{eq:hmax_mean}, most likely scenario detection has advantages in all $t_{obs}$ cases compared to the weighted mean scenario method, as summarized in \figref{fig:hmax_boxplot}.
Furthermore, the weighted mean method does not provide results superior to those of the DTW-based scenario method, even when $t_{obs}$ is set to 15 minutes.
In contrast, the most likely scenario method provides more accurate predictions since its IQRs are smaller than those of the DTW method for the $t_{obs} > 4~\textrm{min}$ condition.
\subsection{Inundation distribution $\bm{h}_{\max}$}
\begin{figure*}[!ht]
    \centering
    \includegraphics[width=\linewidth]{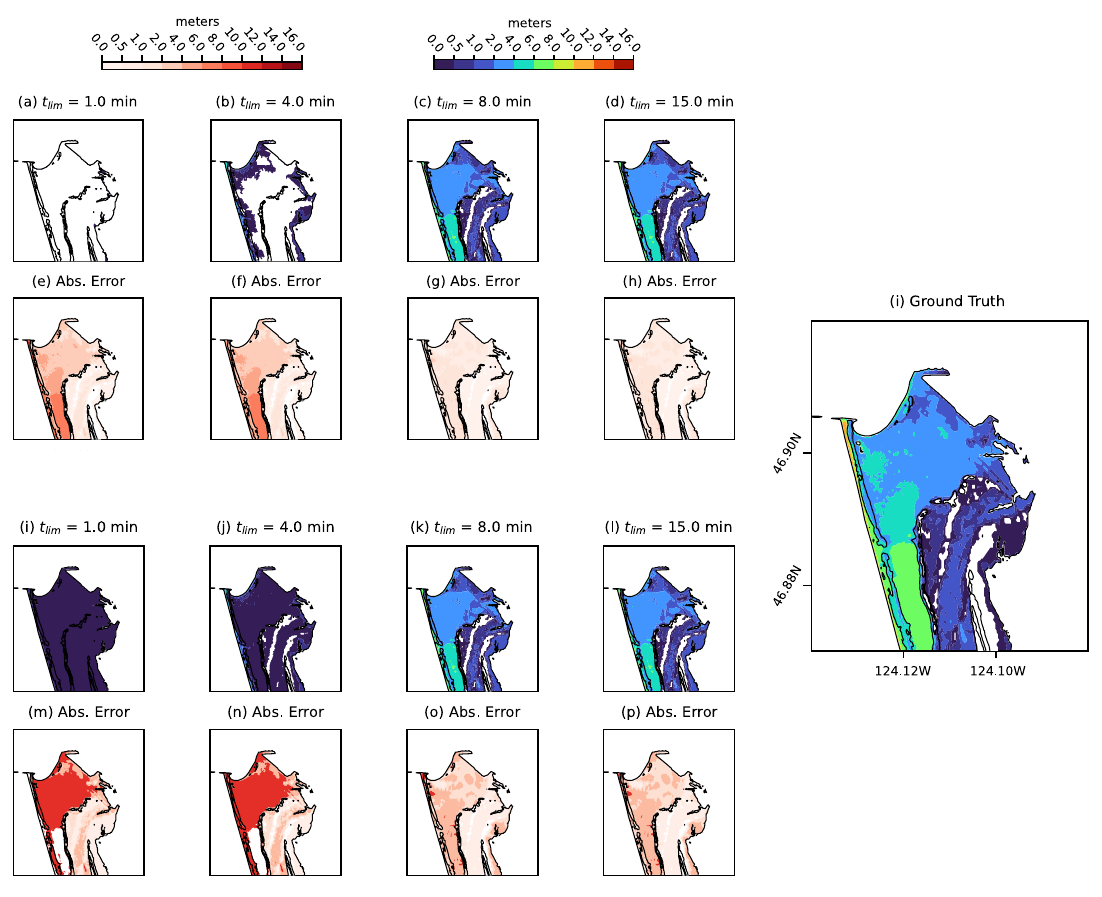}
\vspace{-3 mm}
\caption{The results of inundation depth prediction. (a)-(d) show the inundation depth predictions based on the most likely scenario detection method for observation time windows of $t_{obs}= 1,~4,~8, \textrm{and} 15~$min, and (e)-(h) show the absolute error values. (i)-(l) show the inundation depth predictions based on the weighted mean scenario method for observation time windows of $t_{obs}= 1,~4,~8, \textrm{and} 15~$min, and (e)-(h) show the absolute error values.
The areas with false-negative and false-positive results are illustrated in orange and violet, respectively, and the ground-truth labels are provided in the rightmost panel.}
    \label{fig:comp-inund}
\end{figure*}

\figref{fig:comp-inund} shows the results of inundation prediction based on both methods.
Here, the upper eight panels show the results based on the most likely scenario detection method, and the lower eight panels show those based on the weighted mean scenario method.
The longer the length of the time window is,
the closer the inundation distribution is to the true distribution for both methods
(\figref{fig:comp-inund} (a)-(d), (i)-(l)).
The absolute errors in \figref{fig:comp-inund} (a)-(h) and (m)-(p) decrease in this test case.

\begin{figure*}[!ht]
    \centering
    \includegraphics[width=\linewidth]{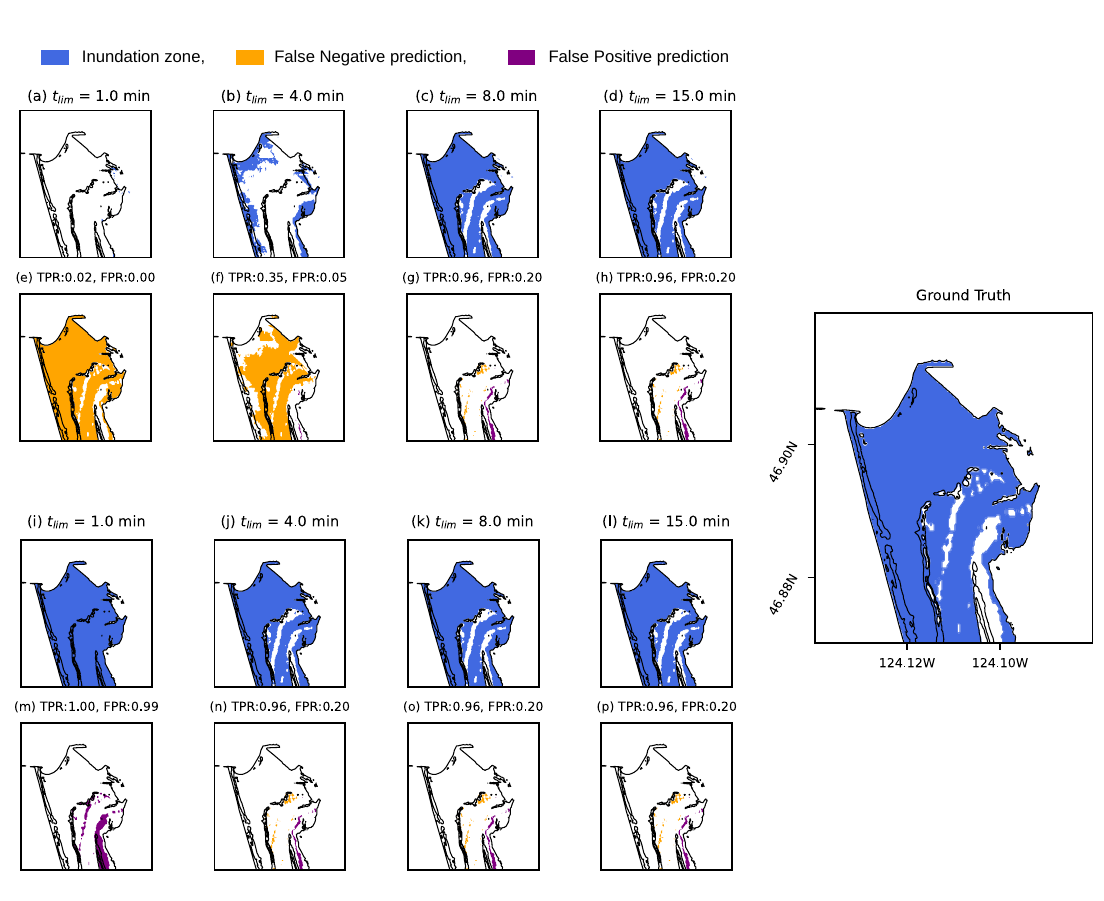}
\vspace{-3 mm}
\caption{The results of inundation area prediction. (a)-(d) show the inundation area (blue) predictions based on the most likely scenario detection method for observation time windows of $t_{obs}= 1,~4,~8, \textrm{and} 15~$min, and (e)-(h) show the areas with erroneous inundation judgments. (i)-(l) show the inundation area predictions based on the weighted mean scenario method for observation time windows of $t_{obs}= 1,~4,~8, \textrm{and} 15~$min, and (m)-(p) show the areas with erroneous inundation judgments.
The areas with false-negative and false-positive results are illustrated in orange and violet, respectively, and the ground-truth labels are provided in the rightmost panel.}
    \label{fig:comp-inund_Bool}
\end{figure*}

To comprehensively evaluate the performance in inundation prediction, we introduce binary evaluation indices.
That is, to confirm the accuracy of the inundation distribution trends, we employ the true-positive rate (TPR) and the false-positive rate (FPR) as evaluation indices:
\begin{equation}
    \textrm{FPR}_{\chi}^{t_{obs}} = \dfrac{n_\textrm{FP}^{t_{obs}}}{n_\textrm{FP}^{t_{obs}} + n_\textrm{TN}^{t_{obs}}}
    \label{eq:false_nr}
\end{equation}
\begin{equation}
    \textrm{TPR}_{\chi}^{t_{obs}} = \dfrac{n_\textrm{TP}^{t_{obs}}}{n_\textrm{TP}^{t_{obs}} + n_\textrm{FN}^{t_{obs}}}
    \label{eq:true_pr}
\end{equation}
Here, $n_\textrm{TP}^{t_{obs}}$, $n_\textrm{FN}^{t_{obs}}$, $n_\textrm{FP}^{t_{obs}}$ and $n_\textrm{TN}$ indicate
\begin{itemize}
\item[$n_\textrm{TP}^{t_{obs}}$]: The number of true-positive predicted grid points at $t_{obs}$(inundated/properly predicted)
\item[$n_\textrm{TN}^{t_{obs}}$]: The number of true-negative predicted grid points at $t_{obs}$(not inundated/properly predicted)
\item[$n_\textrm{FP}^{t_{obs}}$]: The number of false-positive predicted grid points at $t_{obs}$(not inundated/incorrectly predicted)
\item[$n_\textrm{FN}^{t_{obs}}$]: The number of false-negative predicted grid points at $t_{obs}$(inundated/incorrectly predicted)
\end{itemize}
\figref{fig:comp-inund_Bool} shows an example of a false-negative/positive area.
Panels (a)-(e) and (i)-(l) show the predicted inundation zones for each observation time window of $t_{obs} = 1,~4,~8,~10$ min.

In panels (e)-(h) and (m)-(p), the broader yellow zones indicate unpredictable inundation (false-negative prediction).
The points in the violet zones are dry but judged as inundated (false-positive prediction).
The FPR calculated from \eref{eq:false_nr} represents the ratio of the violet zone size to the total number of false predictions and should be close to 0.0 for a good prediction of safe regions.
On the other hand, TPR is expected to be close to 1.0 if the method can predict the actual inundated areas, i.e., the actual hazardous zones.
Specifically, as the ratio of yellow zones in \figref{fig:comp-inund_Bool} decreases, TPR approaches 1.0.

\begin{figure*}[!ht]
    \centering
    \includegraphics[width=0.8\linewidth]{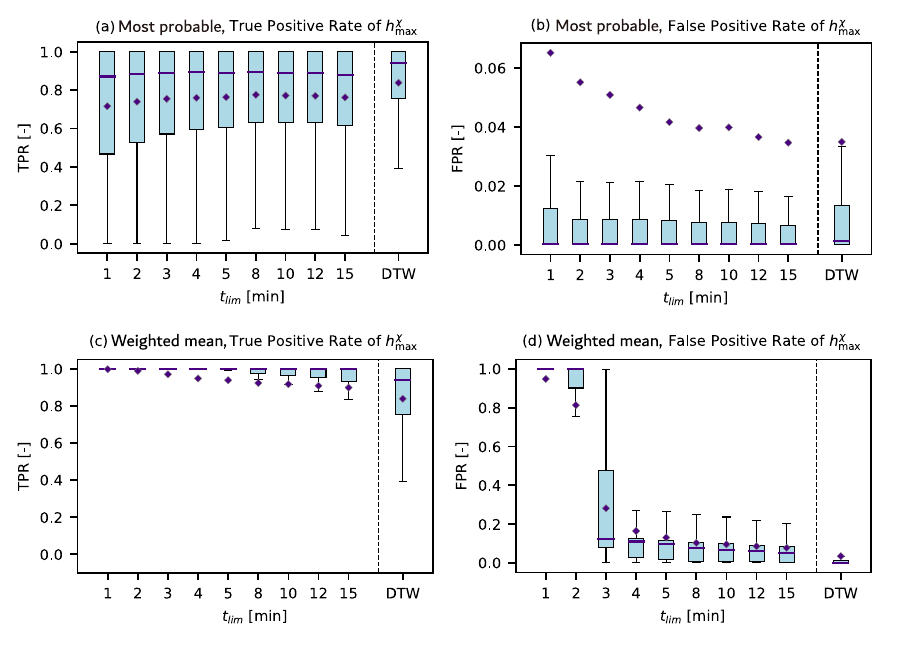}
\vspace{-3 mm}
\caption{The variance in the inundation errors in each observation period $t_{obs}$. (a), (b) Results based on most likely scenario detection, and (c), (d) results based on the weighted mean scenario method. The diamond plot, horizontal line inside the box, and edge lines of the whiskers represent the mean, median, and min/max values, respectively.}
\label{fig:ROC-tlim-boxplo}
\end{figure*}

Here, \figref{fig:ROC-tlim-boxplo} summarizes the TPR and FPR values in all test cases depending on the length of the observation window $t_{obs}$.
Similar to the other figures, this figure provides the results based on the most likely scenario detection method in the upper row and those based on the weighted mean method in the lower row.

For most likely scenario detection, predicting truly inundated zones is somewhat difficult, as shown in \figref{fig:ROC-tlim-boxplo}(a).
The prediction accuracy is not comparable to that of DTW-based scenario detection (rightmost sidebar in (a)) if the longest observation time window $t_{obs}$ is applied.
In contrast, false-negative predictions are avoided in the very early stages of an event, namely, when $t_{obs}=1$ min, as shown in \figref{fig:ROC-tlim-boxplo}(b).
Note that the vertical axis in (b) shows very narrow range (from 0.0 to 0.06) compared to those of the other panels.
According to \figref{fig:ROC-tlim-boxplo}(b), the FPR does not display a large variance and is even superior to the prediction based on the DTW distance (the rightmost sidebar).
The IQRs are all within 0.03 regardless of the length of the observation time window $t_{obs}$.

In contrast, the weighted mean scenario detection method shows its advantages in true inundation zone detection, as shown in \figref{fig:ROC-tlim-boxplo}(c).
However, a longer $t_{obs}$ appears to correspond to increased uncertainty in true-positive prediction.
This finding implies that the number of false-negative predictions $n_\textrm{FN}$ in \eqref{eq:true_pr}, that is, overestimations, increases as the amount of observational data increases.
Another aspect of the weighted mean method is that the FPR variance is greatly reduced if we set $t_{obs} \geq 4.0$, as shown in \figref{fig:ROC-tlim-boxplo}(d), although forecasts that are too early result in erroneous estimates compared to those obtained via the most likely scenario detection method (b).
Another aspect of the weighted mean method is that it cannot provide improved accuracy compared to the shortest DTW distance scenario method.
Based on a comparison of the most likely scenario detection method \eqref{eq:hmax_all_most} and the weighted mean scenario method \eqref{eq:hmax_all_mean}, the weighted mean scenario approach is more appropriate for disaster mitigation purposes.


\section{Discussion}

Thus far, prediction results have been provided and compared in terms of the difference in the scenario chosen or the length of the observation time window.
We now summarize and further discuss the results.

Regarding the evaluation of the maximum wave height, the two methods, namely, the most likely scenario detection method and the weighted mean scenario method, are comparable.
They both predict the maximum wave height with an observation time window $t_{obs}$ of 3-4 minutes.
However, the prediction accuracy tends to not improve after the observation time window reaches such threshold values.

For predicting the maximum onshore inundation depth $H_{\max}^{\chi}$, weighted mean scenario detection is outperformed by the most likely scenario detection and DTW-based scenario detection methods, although its performance improves as the observation time window increases.

With respect to the trends in inundation detection, weighted mean scenario detection is a better choice due to its balanced ability.
Even though the FPR is larger than that obtained using the most likely scenario detection method, overestimations are acceptable compared to underestimations in emergencies, and acquiring accurate TPR values should be prioritized.

In summary, weighted mean scenario detection has advantages over most likely scenario detection when the prediction objective is to determine the inundation trends.
In contrast, most likely scenario detection is preferable if the quantitative amounts or inundation heights of the waves are required.
These findings also suggest that the previously developed framework, namely, most likely scenario detection, is still feasible for a scenario database consisting of diverse fault rupture patterns.
A time window of 3-4 minutes provides acceptable predictions for any purpose with either scenario detection method.
Furthermore, longer observation time windows do not necessarily improve the accuracy in all cases.

Now, we also discuss the reason why the weighted mean method does not provide a significant improvement in prediction. This limitation may arise from the assumption made in the Bayesian updates  \eref{eq:bayesian_update}-\eqref{eq:Mahalanobis}.
Using these equations, we assume that each scenario is considered an independent event. Consequently, the likelihood function $L(E_j)$ is calculated as the finite product:
\begin{equation}
    L(E_j) \approx \prod_{l=1}^{r} \mathcal{N}(\tilde{\alpha}^{(t)}_{l,\chi} \mid \alpha_{l,j}^{(t)}, \sigma_{l}), 
    \label{eq:product}
\end{equation}
where $\mathcal{N}(\bullet | \mu, \sigma)$ expresses the probabilistic density at point $\bullet$ represented by a Gaussian distribution with mean value $\mu$ and standard deviation $\sigma$.
However, when we expect scenario superposition to yield better predictions, \eref{eq:product} is not valid because each scenario should not be considered independent. Instead, scenarios are assumed to occur simultaneously. In such scenarios, the probability values obtained from sequential updates become ``weights'', characterizing the effectiveness of each scenario in explaining the current event.

\section{Conclusions}
This study aimed to elucidate the performance of a previous tsunami scenario detection framework with diverse tsunami scenario databases consisting of complex fault ruptures with heterogeneous slip distributions.
Specifically, the effectiveness of scenario superposition was investigated in terms of its advantages over the previous most likely scenario detection method.
Additionally, how the length of the observation time window affects the accuracy of both methods was determined.
An existing 1771 tsunami scenario database, which comprises synthetic wave height records and inundation distributions, was used for this purpose.
The complex structure of the database, which was assumed to be due to the heterogeneous patterns of slips, was identified, and the scenario diversities were inferred.
From the perspectives of the tsunami arrival time and the maximum inundation depth, we established several candidates for observational time windows shorter than 15 minutes and split the database into five testing and training sets.

From the results, we draw the following conclusions:
\begin{itemize}
\item The weighted mean scenario method does not necessarily provide improved accuracy compared to the most probable scenario detection method. \\
 \item  The weighted mean scenario method slightly better predicted the inundation distribution compared to the most probable scenario detection method.\\
\item  The length of the observation time window does not strongly influence the prediction accuracy if the window length is more than 3-4 minutes. \\
\item The abovementioned tendency was common between the two methods. \\
\item The sequential probability update process should be reformulated for the purpose of scenario superposition.
\end{itemize}

From these findings, we conclude that the current framework performs well for a scenario database comprising diverse fault rupture patterns to a certain extent, but there is room for improvement in achieving better prediction results through scenario superposition.
To do so, a new formulation for the probability update process should be established.
Additionally, addressing challenges associated with shorter observation time windows, particularly those within 3 minutes, could enhance the effectiveness of tsunami evacuation efforts.

\acknowledgements
This work was supported by a JSPS Grant-in-Aid for Young Scientists (Start-up) JP21K20441 and by the Grants for Operations and the Core Research Cluster of Disaster Science, Tohoku University.
The authors would like to acknowledge Prof. Daniel Abramson, Prof. Ann Bostrom, Prof. Liz Maly, Dr. Robert Baraldi and Christopher M. Liu for generously offering their time to discuss the contents of this work.

 \bibliographystyle{jdrbibtex}
 \bibliography{main.bib}

\begin{thebibliography}{10}

\bibitem{sugawara2020trigger}
D.~Sugawara, ``Chapter 4 - Trigger mechanisms and hydrodynamics of tsunamis'',
  Geological Records of Tsunamis and Other Extreme Waves, pp.~ 47--73, 2020.
\newblock doi:10.1016/B978-0-12-815686-5.00004-3.

\bibitem{cisternas2005predecessors}
M.~Cisternas, B.~F. Atwater, F.~Torrej{\'o}n, Y.~Sawai, G.~Machuca, M.~Lagos,
  A.~Eipert, C.~Youlton, I.~Salgado, T.~Kamataki, et~al., ``Predecessors of the
  giant 1960 Chile earthquake'', Nature, Vol.437, No.7057, pp.~404--407, 2005.

\bibitem{Franco_2010jdr}
G.~Franco and W.~Siembieda, ``Chile’s 2010 M8.8 Earthquake and Tsunami:
  Initial Observations on Resilience'', Journal of Disaster Research, Vol.5,
  No.5, pp.~577--590, 2010.
\newblock doi:10.20965/jdr.2010.p0577.

\bibitem{lay2005great}
T.~Lay, H.~Kanamori, C.~J. Ammon, M.~Nettles, S.~N. Ward, R.~C. Aster, S.~L.
  Beck, S.~L. Bilek, M.~R. Brudzinski, R.~Butler, et~al., ``The great
  Sumatra-Andaman earthquake of 26 december 2004'', science, Vol.308, No.5725,
  pp.~1127--1133, 2005.

\bibitem{Tsuji_2006jdr}
Y.~Tsuji, Y.~Tanioka, H.~Matsutomi, Y.~Nishimura, T.~Kamataki, Y.~Murakami,
  T.~Sakakiyama, A.~Moore, G.~Gelfenbaum, S.~Nugroho, B.~Waluyo, I.~Sukanta,
  R.~Triyono, and Y.~Namegaya, ``Damage and Height Distribution of Sumatra
  Earthquake-Tsunami of December 26, 2004, in Banda Aceh City and its
  Environs'', Journal of Disaster Research, Vol.1, No.1, pp.~103--115, 2006.
\newblock doi:10.20965/jdr.2006.p0103.

\bibitem{Takahashi_2017jdr}
N.~Takahashi, K.~Imai, M.~Ishibashi, K.~Sueki, R.~Obayashi, T.~Tanabe,
  F.~Tamazawa, T.~Baba, and Y.~Kaneda, ``Real-Time Tsunami Prediction System
  Using DONET'', Journal of Disaster Research, Vol.12, No.4, pp.~766--774,
  2017.
\newblock doi:10.20965/jdr.2017.p0766.

\bibitem{Matsumoto_2017jdr}
H.~Matsumoto, M.~A. Nosov, S.~V. Kolesov, and Y.~Kaneda, ``Analysis of Pressure
  and Acceleration Signals from the 2011 Tohoku Earthquake Observed by the
  DONET Seafloor Network'', Journal of Disaster Research, Vol.12, No.1,
  pp.~163--175, 2017.
\newblock doi:10.20965/jdr.2017.p0163.

\bibitem{Aoi_2019jdr}
S.~Aoi, W.~Suzuki, N.~Y. Chikasada, T.~Miyoshi, T.~Arikawa, and K.~Seki,
  ``Development and Utilization of Real-Time Tsunami Inundation Forecast System
  Using S-net Data'', Journal of Disaster Research, Vol.14, No.2, pp.~212--224,
  2019.
\newblock doi:10.20965/jdr.2019.p0212.

\bibitem{DART}
N.~Oceanic and A.~Administration.
\newblock ``Deep-Ocean Assessment and Reporting of Tsunamis (DART(R))'', 2005.
\newblock [accessed 07.09.2022].
\newblock doi:10.7289/V5F18WNS.

\bibitem{bernard2015evolution}
E.~Bernard and V.~Titov, ``Evolution of tsunami warning systems and products'',
  Philosophical Transactions of the Royal Society A: Mathematical, Physical and
  Engineering Sciences, Vol.373, No.2053, pp.~20140371, 2015.

\bibitem{titov2005real}
V.~V. Titov, F.~I. Gonzalez, E.~Bernard, M.~C. Eble, H.~O. Mofjeld, J.~C.
  Newman, and A.~J. Venturato, ``Real-time tsunami forecasting: Challenges and
  solutions'', Natural Hazards, Vol.35, pp.~35--41, 2005.

\bibitem{fauzi2019machine}
A.~Fauzi and N.~Mizutani, ``Machine learning algorithms for real-time tsunami
  inundation forecasting: a case study in \textsc{N}ankai region'', Pure and
  Applied Geophysics, pp.~ 1--14, 2019.

\bibitem{liu2021comparison}
C.~M. Liu, D.~Rim, R.~Baraldi, and R.~J. LeVeque, ``Comparison of machine
  learning approaches for tsunami forecasting from sparse observations'', Pure
  and Applied Geophysics, Vol.178, No.12, pp.~5129--5153, 2021.

\bibitem{mulia2022machine}
I.~E. Mulia, N.~Ueda, T.~Miyoshi, A.~R. Gusman, and K.~Satake, ``Machine
  learning-based tsunami inundation prediction derived from offshore
  observations'', Nature Communications, Vol.13, No.1, pp.~5489, 2022.

\bibitem{kamiya2022numerical}
M.~Kamiya, Y.~Igarashi, M.~Okada, and T.~Baba, ``Numerical experiments on
  tsunami flow depth prediction for clustered areas using regression and
  machine learning models'', Earth, Planets and Space, Vol.74, No.1, pp.~127,
  2022.

\bibitem{nomura2022sequential}
R.~Nomura, S.~Fujita, J.~M. Galbreath, Y.~Otake, S.~Moriguchi, S.~Koshimura,
  R.~J. LeVeque, and K.~Terada, ``Sequential Bayesian Update to Detect the Most
  Likely Tsunami Scenario Using Observational Wave Sequences'', Journal of
  Geophysical Research: Oceans, Vol.127, No.10, pp.~e2021JC018324, 2022.

\bibitem{geist2002complex}
E.~L. Geist, ``Complex earthquake rupture and local tsunamis'', Journal of
  Geophysical Research: Solid Earth, Vol.107, No.B5, pp.~ESE--2, 2002.

\bibitem{koshimura_2022_zenodo}
S.~Koshimura and R.~Nomura.
\newblock ``{Data from 666 earthquake/tsunami scenario simulations targeting
  Nankai subduction}'', Jul. 2022.
\newblock \url{https://doi.org/10.5281/zenodo.6785643},
  doi:10.5281/zenodo.6785643.

\bibitem{williamson2020source}
A.~L. Williamson, D.~Rim, L.~M. Adams, R.~J. LeVeque, D.~Melgar, and F.~I.
  Gonz{\'a}lez, ``A source clustering approach for efficient inundation
  modeling and regional scale probabilistic tsunami hazard assessment'',
  Frontiers in Earth Science, pp.~ 442, 2020.

\bibitem{fujita2024optimization}
S.~Fujita, R.~Nomura, S.~Moriguchi, Y.~Otake, S.~Koshimura, R.~J. LeVeque, and
  K.~Terada, ``Optimization of a tsunami gauge configuration for
  pseudo-super-resolution of wave height distribution'', Earth and Space
  Science, Vol.11, No.2, pp.~e2023EA003144, 2024.

\bibitem{liang2002proper}
Y.~Liang, H.~Lee, S.~Lim, W.~Lin, K.~Lee, and C.~Wu, ``Proper orthogonal
  decomposition and its applications--Part I: Theory'', Journal of Sound and
  vibration, Vol.252, No.3, pp.~527--544, 2002.

\bibitem{kerschen2005method}
G.~Kerschen, J.-c. Golinval, A.~F. Vakakis, and L.~A. Bergman, ``The method of
  proper orthogonal decomposition for dynamical characterization and order
  reduction of mechanical systems: an overview'', Nonlinear dynamics, Vol.41,
  No.1, pp.~147--169, 2005.

\bibitem{yamamoto2016multi}
N.~Yamamoto, S.~Aoi, K.~Hirata, W.~Suzuki, T.~Kunugi, and H.~Nakamura,
  ``Multi-index method using offshore ocean-bottom pressure data for real-time
  tsunami forecast'', Earth, Planets and Space, Vol.68, pp.~1--14, 2016.

\bibitem{bellman1959adaptive}
R.~Bellman and R.~Kalaba, ``On adaptive control processes'', IRE Transactions
  on Automatic Control, Vol.4, No.2, pp.~1--9, 1959.

\bibitem{senin2008dynamic}
P.~Senin, ``Dynamic time warping algorithm review'', Information and Computer
  Science Department University of Hawaii at Manoa Honolulu, USA, Vol.855,
  No.1-23, pp.~40, 2008.

\bibitem{lee2020clustering}
S.~Lee, J.~Kim, J.~Hwang, E.~Lee, K.-J. Lee, J.~Oh, J.~Park, and T.-Y. Heo,
  ``Clustering of time series water quality data using dynamic time warping: A
  case study from the Bukhan River water quality monitoring network'', Water,
  Vol.12, No.9, pp.~2411, 2020.

\bibitem{ouyang2010similarity}
R.~Ouyang, L.~Ren, W.~Cheng, and C.~Zhou, ``Similarity search and pattern
  discovery in hydrological time series data mining'', Hydrological Processes:
  An International Journal, Vol.24, No.9, pp.~1198--1210, 2010.

\bibitem{thomas1987earthquake}
T.~H. Heaton and S.~H. Hartzell, ``Earthquake Hazards on the Cascadia
  Subduction Zone'', Science, Vol.236, No.4798, pp.~162--168, 1987.
\newblock doi:10.1126/science.236.4798.162.

\bibitem{atwater2011orphan}
B.~F. Atwater, S.~Musumi-Rokkaku, K.~Satake, Y.~Tsuji, and D.~K. Yamaguchi,
  ``The orphan tsunami of 1700: Japanese clues to a parent earthquake in North
  America'', University of Washington Press, 2011.

\bibitem{goldfinger2003holocene}
C.~Goldfinger, C.~H. Nelson, J.~E. Johnson, and S.~S. Party, ``Holocene
  earthquake records from the Cascadia subduction zone and northern San Andreas
  fault based on precise dating of offshore turbidites'', Annual Review of
  Earth and Planetary Sciences, Vol.31, No.1, pp.~555--577, 2003.

\bibitem{leveque2016generating}
R.~J. LeVeque, K.~Waagan, F.~I. Gonz{\'a}lez, D.~Rim, and G.~Lin, ``Generating
  random earthquake events for probabilistic tsunami hazard assessment'', In
  Global Tsunami Science: Past and Future, Volume I, pp.~ 3671--3692, Springer,
  2016.

\bibitem{melgar2016kinematic}
D.~Melgar, R.~J. LeVeque, D.~S. Dreger, and R.~M. Allen, ``Kinematic rupture
  scenarios and synthetic displacement data: An example application to the
  Cascadia subduction zone'', Journal of Geophysical Research: Solid Earth,
  Vol.121, No.9, pp.~6658--6674, 2016.

\bibitem{melgar2021mudpy}
D.~Melgar, T.~Lin, Q.~Kong, christineruhl, and B.~Marfito.
\newblock ``dmelgarm/MudPy: v1.3'', Sep. 2021.
\newblock \url{https://doi.org/10.5281/zenodo.5397091},
  doi:10.5281/zenodo.5397091.

\bibitem{koshimura_2023_zenodo}
S.~Koshimura and S.~Fujita.
\newblock ``{Data from 1564 earthquake/tsunami scenario simulations targeting
  the Nankai Trough subduction zone}'', Aug. 2023.
\newblock \url{https://doi.org/10.5281/zenodo.8287917},
  doi:10.5281/zenodo.8287917.

\bibitem{clawpack}
{Clawpack Development Team}.
\newblock ``Clawpack software'', 2020.
\newblock Version 5.7.1.
\newblock \url{http://www.clawpack.org},
  doi:https://doi.org/10.5281/zenodo.4025432.

\bibitem{berger2011geoclaw}
M.~J. Berger, D.~L. George, R.~J. LeVeque, and K.~T. Mandli, ``The {GeoClaw}
  software for depth-averaged flows with adaptive refinement'', Adv. Water
  Res., Vol.34, pp.~1195--1206, 2011.
\newblock \url{www.clawpack.org/links/papers/awr11}.

\bibitem{mandli2016clawpack}
K.~T. Mandli, A.~J. Ahmadia, M.~Berger, D.~Calhoun, D.~L. George,
  Y.~Hadjimichael, D.~I. Ketcheson, G.~I. Lemoine, and R.~J. LeVeque,
  ``Clawpack: building an open source ecosystem for solving hyperbolic PDEs'',
  PeerJ Computer Science, Vol.2, pp.~e68, 2016.
\newblock doi:10.7717/peerj-cs.68.

\bibitem{UW_repo}
``Archive of simulation data and results for Frontiers 2020 paper''.
\newblock accessed 04.22.2024.
\newblock \url{http://depts.washington.edu/ptha/frontiers2020a/}.

\end{thebibliography}


\appendix
\section{Appendix: Evaluation based on the shortest DTW distance scenario method}
\label{app:DTW_prediction}
As we reported in \secref{sec:3method}, \ref{sec:results}, we employed the DTW method as the benchmark for validating the reasonability of Bayesian scenario detection method.
As in the other two methods, we established 10 different observation time windows, $t_{obs} =
1, 2, 3, 4, 5, 8, 10, 12, 15$ min and $4 \textrm{hr}$.

In each observation time window, we calculated the DTW distance based on wave height data.

\begin{figure*}
    \centering
    \includegraphics[width=0.8\linewidth]{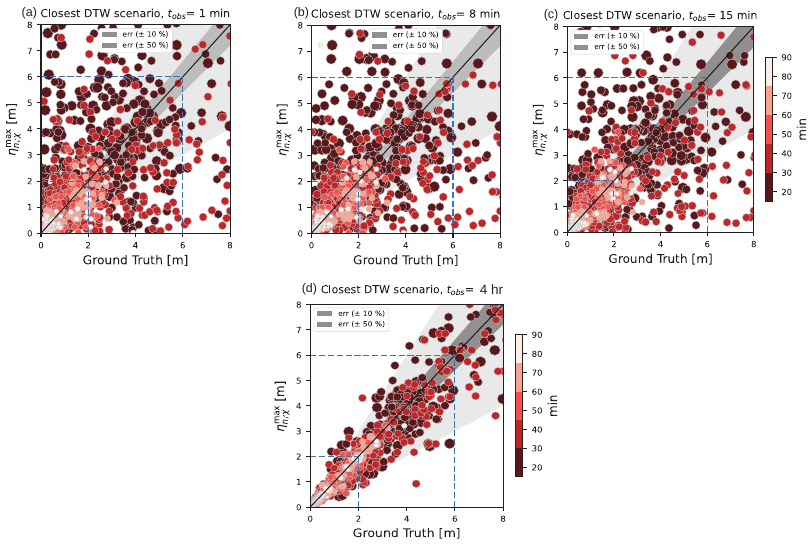}
\caption{The results of maximum wave height $\eta_{n',\chi}^{\max}$ detection based on dynamic time warping (DTW). (a), (b), and (c) show the predictions based on the shortest DTW distance scenario method for observation time windows of $t_{obs}$ =1~min, 8~min, and 15~min, respectively. (d) shows the results obtained with all the wave history data ($t_{obs}$=4 hr) corresponding to the rightmost box in \figref{fig:etamax_sigma_box}, \ref{fig:hmax_boxplot}, \ref{fig:ROC-tlim-boxplo}.}
    \label{fig:etamax_DTWscenario}
\end{figure*}

Here, \figref{fig:etamax_DTWscenario} shows the maximum wave heights obtained based on the shortest DTW distances.
Although good prediction results are obtained if we use the entire wave history dataset, as shown in \figref{fig:etamax_DTWscenario}(d), the shortest DTW distance scenario method does not provide a good prediction if the observation time window $t_{obs}$ is less than 15 minutes, as shown in panels (a)-(c).

\begin{figure}
    \centering
    \includegraphics[width=0.8\linewidth]{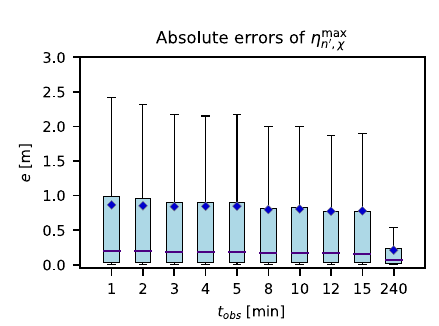}
\caption{The variance in the inundation errors in each observation period $t_{obs}$ based on the shortest DTW distance scenario method. The rightmost box shows the results obtained with all the wave history data ($t_{obs}$=4 hr) and corresponds to the rightmost box in \figref{fig:etamax_sigma_box}, \ref{fig:hmax_boxplot}, and \ref{fig:ROC-tlim-boxplo}.}
    \label{fig:etamax-tlim-boxplot_dtw}
\end{figure}

This tendency is also confirmed by the box plot in \figref{fig:etamax-tlim-boxplot_dtw}, which summarizes the absolute errors of the maximum wave heights.
These results and comparisons with those of the other two methods are presented in \secref{sec:results}.
Bayesian update-based probabilistic evaluation has an advantage over methods that use standard time-series features.


\vspace{25pt}


\end{document}